\documentclass{emulateapj}
\usepackage{apjfonts}

\shortauthors{Labb\'e et al.}
\shorttitle{IRAC observations of red $z>2$ galaxies}
\slugcomment{}

\begin{document}

\def\figa{
\leavevmode
\vspace*{0.5cm}
 \hbox{\hspace{-0.1cm} \includegraphics[bb=42 113 365 359,width=0.52\textwidth]{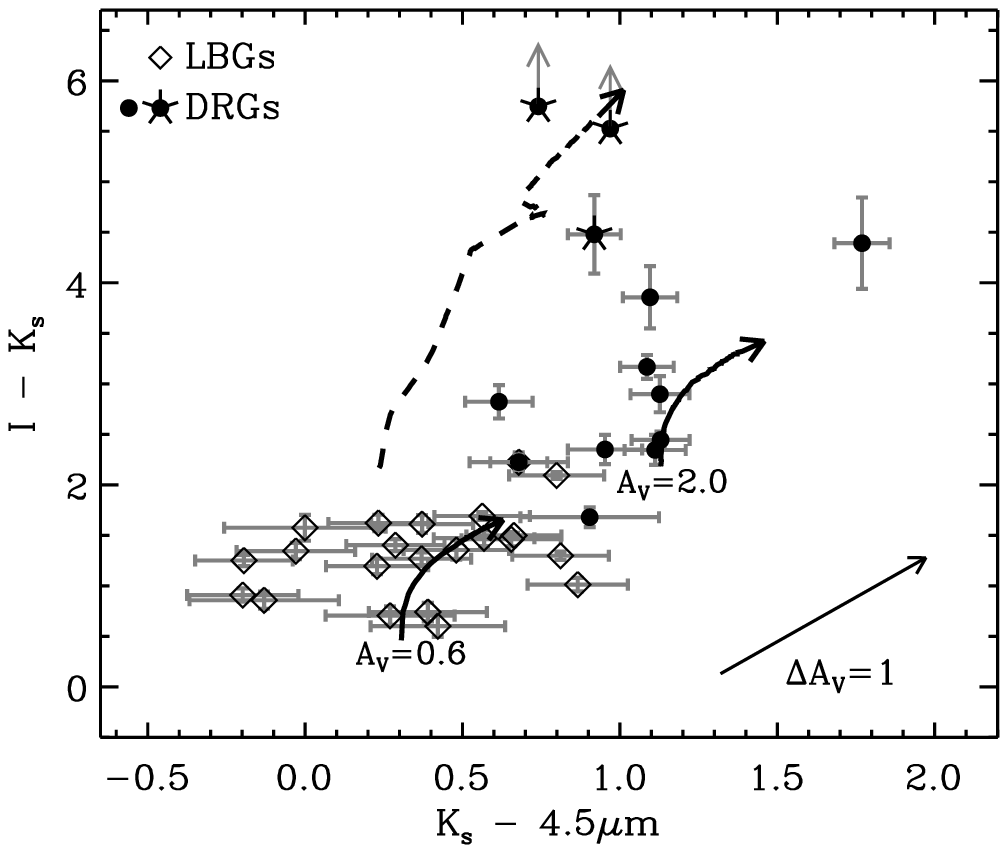} }
\figcaption{Observed $I - K_s$ versus $K_s - 4.5\mu$m color-color 
diagram of two samples of $z>2$ galaxies. Distant Red Galaxies 
(DRGs; {\it filled circles and stars}), occupy a different color 
region than $z\sim2.5$  Lyman Break Galaxies (LBGs; {\it diamonds}). DRGs 
that are best described by an old single-age burst 
model (SSP; {\it stars}) have colors distinct from those better 
described by constant star-forming models (CSF) and dust, 
indicating that IRAC fluxes may help
in separating these populations. The curves show 
the color-evolution tracks of Bruzual \& Charlot (2003)
models at a fixed $z=2.6$: an SSP model with ages ranging
from 0.3 to 3 Gyr ({\it dashed line}) and two CSF
models with ages ranging from 0.1 to 3 Gyr, and reddenings 
of $A_V=0.6$ and $A_V=2.0$ respectively ({\it solid lines}). 
The vector indicates a reddening of $A_V=1$ mag for a 
Calzetti et al. (2000) law. \label{fig.a}}
}

\def\figb{
\begin{figure*}[t]  
\hbox{ \hspace{-0.25cm} \includegraphics[width=1.02\textwidth]{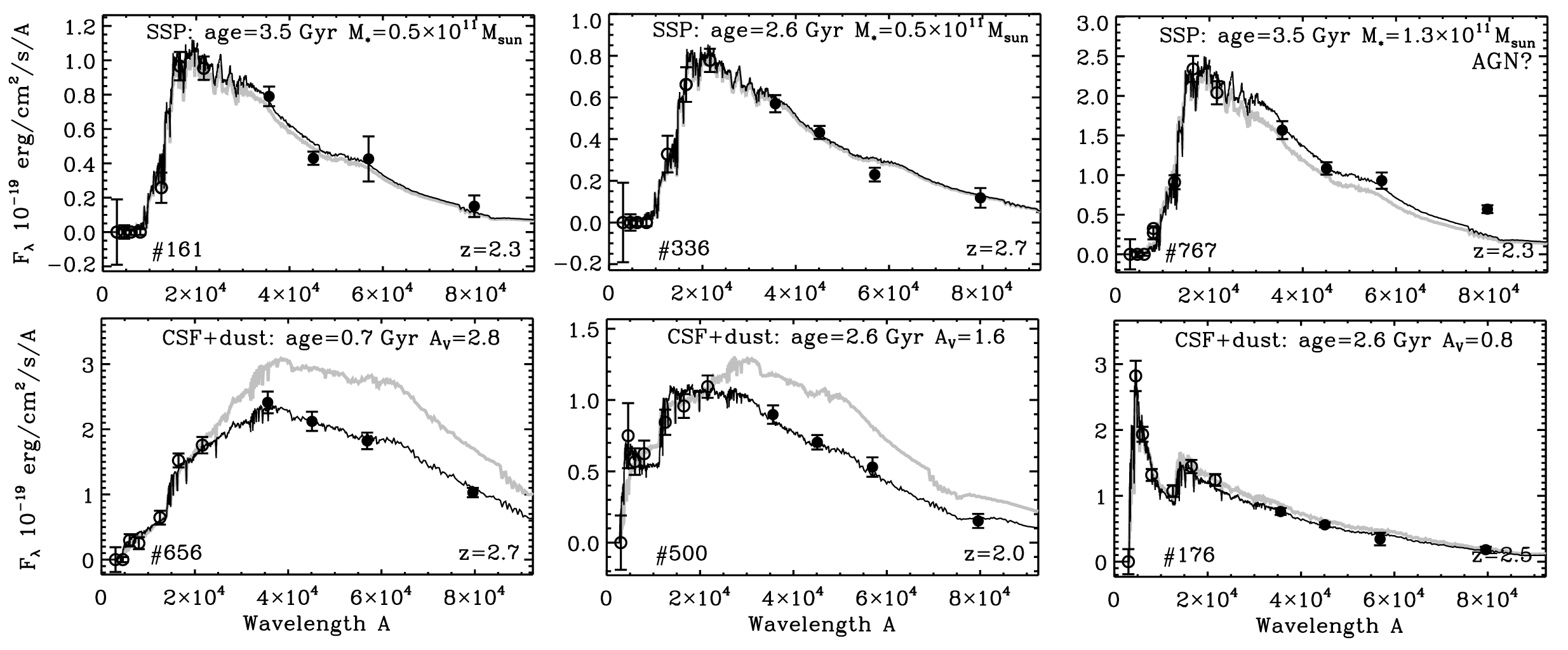} }
\figcaption{The UV-to-$8\mu$m spectral energy distributions (SEDs) of 
Distant Red Galaxies (DRGs). The best-fit Bruzual \& Charlot (2003)
models to the full SEDs ({\it black solid lines}) are shown, but also 
predictions based on the ultradeep optical and near-infrared fluxes only 
({\it gray solid lines}). {\it Top row:} the 3 DRGs best fit with 
an single-age burst (SSP) model. The mid-infrared (MIR) 
IRAC fluxes ({\it filled circles}) directly confirm the predictions.
Old and dead DRGs have bluer $K_s-4.5\mu$m colors than 
very dusty star-forming DRGs. Galaxy 767 shows 
a flux excess at $8\mu$m, possibly related to AGN activity. {\it Bottom row:} the 
MIR predictions for galaxies better fit with constant
star forming  models and Calzetti et al. (2000) dust-reddening.
For highly-reddened galaxies, the MIR observations can be 
very different from the predictions. Here IRAC fluxes help to
better constrain the total age, dust content, and stellar mass 
in the models. \label{fig.b}}
\end{figure*}
}

\def\figc{
\begin{figure*}[t]
\hbox { \hspace{-0.25cm} \includegraphics[width=1.02\textwidth]{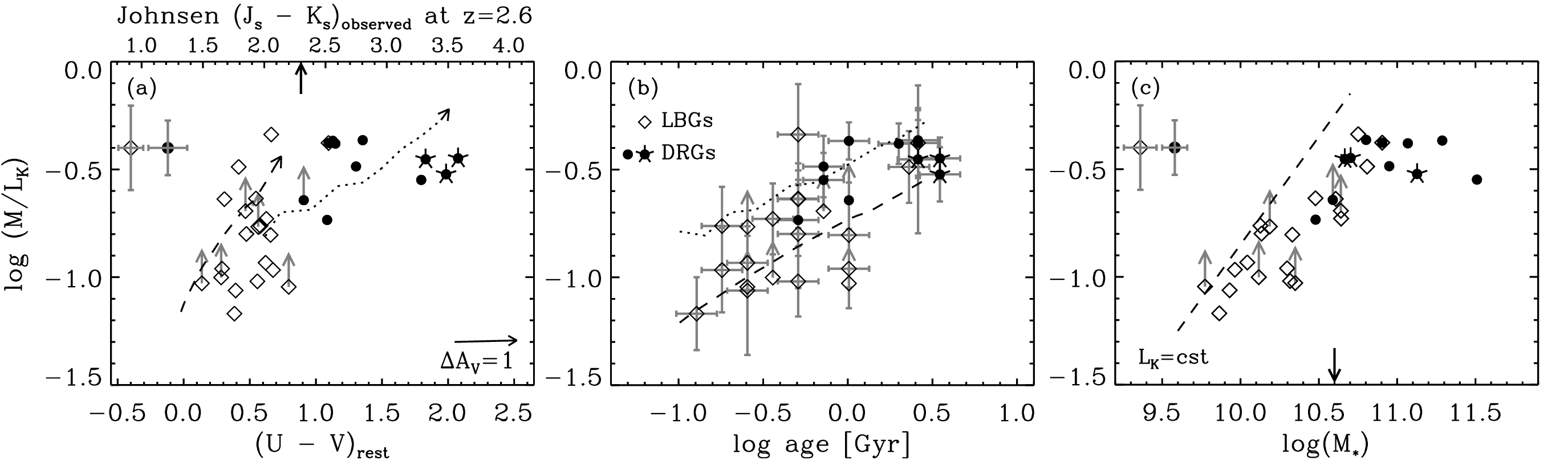} }
\figcaption{The mass-to-light ratios in the rest-frame $K-$band ($M/L_K$) of 
distant red galaxies (DRGs) and $z=2-3$ Lyman break galaxies (LBGs).
Symbols are as in Fig.~1. ({\it a}) $M/L_K$ versus rest-frame $U-V$ color, showing a clear correlation. 
On top we indicate the corresponding observed $J_s-K_s$ color and the $J_s-K_s>2.3$ limit ({\it upward arrow}). 
Color tracks of an SSP ({\it dotted line}) and CSF model ({\it dashed line}) are shown.
({\it b}) The relation between $M/L_K$ and best-fit age. The 
$M/L_K$ is more sensitive to the age of the stellar population than to dust extinction,
hence the DRGs are on average older, not just more dusty versions of LBGs. ({\it c}) 
$M/L_K$ versus stellar mass. The highest-mass galaxies all have high $M/L_K$ ratios. 
The mass completeness limit ({\it downward arrow}) corresponds to our observed 
$K_{s,tot}=22.5$ magnitude cut-off and a maximal $M/L$.  A selection by rest-frame 
K-light corresponds to taking everything to the right of the 
constant $L_K$ line (\textit{dashed line}).  Obviously this is still far
from a vertical cut-off which would represent a selection by stellar mass.
}
\end{figure*}
}

\title{IRAC Mid-Infrared Imaging of the Hubble Deep Field South:
 Star Formation Histories and Stellar Masses of Red Galaxies at $z>2$\altaffilmark{1}}
\author{
Ivo Labb\'{e}\altaffilmark{2}, Jiasheng Huang\altaffilmark{3},
Marijn Franx\altaffilmark{4}, Gregory Rudnick\altaffilmark{5},  
Pauline Barmby\altaffilmark{3}, Emanuele Daddi\altaffilmark{6}, Pieter G. van Dokkum\altaffilmark{7}, 
Giovanni G. Fazio\altaffilmark{3}, Natascha M. F\"{o}rster Schreiber\altaffilmark{8}, 
Alan F. M. Moorwood\altaffilmark{9}, Hans-Walter Rix\altaffilmark{10}, Huub R\"{o}ttgering\altaffilmark{4}, 
Ignacio Trujillo\altaffilmark{10}, Paul van der Werf\altaffilmark{4}}

\altaffiltext{1}{Based on observations with the {\em Spitzer Space Telescope}, which is 
operated by the Jet Propulsion Laboratory, California Institute of Technology under 
NASA contract 1407.  Support for this work was provided by NASA through contract 
125790 issued by JPL/Caltech.  Based on service mode observations collected at
the European Southern Observatory, Paranal, Chile
(LP Program 164.O-0612). Based on observations with the
NASA/ESA {\em Hubble Space Telescope}, obtained at the Space
Telescope Science Institute which is operated by AURA, Inc.,
under NASA contract NAS5-26555
}

\accepted{for publication in the ApJ Letters}

\altaffiltext{2}{Carnegie fellow, OCIW, 813 Santa Barbara Street, Pasadena, CA 91101 [e-mail: {\tt ivo@ociw.edu}]}
\altaffiltext{3}{CFA, 60 Garden Street Cambridge, MA 02138}
\altaffiltext{4}{Leiden Observatory, P.O. Box 9513, NL-2300 RA,
Leiden, The Netherlands}
\altaffiltext{5}{Goldberg fellow, NOAO, 950 N. Cherry Ave, Tucson, AZ 85719}
\altaffiltext{6}{Spitzer fellow, NOAO, 950 N. Cherry Ave, Tucson, AZ 85719}
\altaffiltext{7}{Department of Astronomy, Yale University, P.O. Box 208101, New Haven, CT 06520-8101}
\altaffiltext{8}{MPE, Giessenbackstrasse, D-85748, Garching, Germany}
\altaffiltext{9}{ESO, D-85748, Garching, Germany}
\altaffiltext{10}{MPIA, K\"onigstuhl 17, D-69117, Heidelberg, Germany}
 
\begin{abstract}
We present deep $3.6-8\mu$m imaging of the Hubble Deep Field South
with IRAC on the {\it Spitzer Space Telescope}.
We study Distant Red Galaxies (DRGs) at $z>2$ selected by
$J_s-K_s>2.3$ and compare them to a sample of Lyman Break Galaxies
(LBGs) at $z=2-3$.  
The observed UV-to-$8\mu$m spectral energy distributions are fit with stellar
population models to constrain star formation histories and  derive
stellar masses.  We find that 70\% of the DRGs are best described by
dust-reddened  star forming models and 30\% are very well fit with old 
and ``dead'' models. Using only the $I-K_s$ and $K_s-4.5\mu$m
colors we can effectively separate the two groups.
The dead systems are among the most massive at $z\sim2.5$ 
(mean stellar mass $<M_*> = 0.8 \times 10^{11} M_{\sun}$) 
and likely formed most of their stellar mass at $z>5$. 
To a limit of $0.5\times 10^{11} M_{\sun}$ their number density is 
 $\sim10\times$ lower than that of local early-type galaxies. 
Furthermore, we use the IRAC photometry to derive rest-frame 
near-infrared $J, H,$ and $K$ fluxes.
The DRGs and LBGs together show a large variation (a factor of 6) in the 
rest-frame $K-$band mass-to-light ratios ($M/L_K$), implying that even a 
Spitzer $8\mu$m$-$selected sample would be very different from a 
mass-selected sample. The average $M/L_K$ of the DRGs is about three times
higher than that of the LBGs, and DRGs dominate the high-mass end.
The $M/L_K$ ratios and ages of the two samples appear to correlate with 
derived stellar mass, with the most massive galaxies being the oldest
and having the highest mass-to-light ratios, similar as found 
in the low-redshift universe. %
\end{abstract}


\clearpage

\section{Introduction}
Galaxies at $z>2$ exhibit very diverse properties: they range from the
blue Lyman Break Galaxies (LBGs) which are bright in the rest-frame ultraviolet
(Steidel et al 1996a,b) to the Distant Red Galaxies (DRGs)
which are generally faint in the rest-frame UV and have fairly red
rest-frame optical colors. The DRGs are selected in the observers'
near-infrared (NIR) by the simple criterion $J_s - K_s > 2.3$ (Franx et al. 2003,
van Dokkum et al. 2003). 
The variety in the galaxy population at $z>2$ is comparable to that
seen in the local universe, where colors range from very blue for
young starbursting galaxies to very red for old elliptical galaxies.
An urgent question is what causes the red colors of DRGs
at $z>2$. Are they ``old and dead'', or are they actively star forming 
and more dusty than U-dropout galaxies?

First analyses of the
optical-to-NIR Spectral Energy Distributions (SEDs) and 
spectra have suggested that both effects play a 
role: they have higher ages, contain more dust, and have higher 
mass-to-light ($M/L$) ratios in the rest-frame optical than LBGs (Franx 
et al. 2003, van Dokkum et al. 2004, Forster Schreiber et al. 2004; F04).
Furthermore, many have high star formation rates 
$> 100~M_{\sun}$yr$^{-1}$ (van Dokkum et al. 2004, F04).
Unfortunately, the number of DRGs with rest-frame optical spectroscopy
is very small. Inferences inevitably depend on
SED fitting, which has large uncertainties \citep[see e.g.,][]{PDF01}.
The SED constraints on the stellar and dust content are expected
to improve significantly by extending the photometry to 
the rest-frame near-infrared.

Here we present the first results on rest-frame NIR photometry of 
DRGs and LBGs in the Hubble Deep Field South (HDFS) as observed with the 
Infrared Array Camera (IRAC; Fazio et al. 2004) on the {\it Spitzer Space Telescope}.
Where necessary, we assume an $\Omega_M=0.3, \Omega_\Lambda=0.7,$ 
cosmology with $H_0=70$~km~s$^{-1}$Mpc$^{-1}$, and a Salpeter 
Initial Mass Function (IMF) between 0.1 and $100~M_\sun$. 
Magnitudes are expressed in the AB photometric system 
unless explicitly stated otherwise.

\section{The Observations, Photometry, and Sample Selection}
We have observed the 5 arcmin$^2$ HDFS/WFPC2 field with 
the IRAC camera integrating 1 hour each in the mid-infrared 
(MIR) 3.6, 4.5, 5.8, and 8$\mu$m$-$bands. The observations, data reduction, 
and photometry will be described by Labb\'e et al. (in preparation). 
Briefly, we reduced and calibrated the data using standard procedures 
\cite[e.g.,][]{Ba04}. The limiting depths at 
3.6, 4.5, 5.8, and 8$\mu$m are 25.0, 24.6, 22.6, and 22.5 mag, respectively  
(5$\sigma$, 3$\arcsec$ diameter aperture). We PSF-matched 
the IRAC images and a very deep $K_s-$image (Labb\'e et al. 2003) to the 
$8\mu$m$-$band. Nearby sources were fitted and subtracted 
to avoid confusion (see Labb\'e et al. 2004). We measured fluxes 
in $4\farcs4$ diameter apertures. The $K_s-IRAC$ colors were then 
combined with the HDFS-catalog of Labb\'e et al. (2003) resulting 
in 11-band photometry from 0.3 to 8$\mu$m. In addition to the 
photometric errors, we add in quadrature an error of 10\% to reflect 
absolute calibration uncertainties of IRAC. 

We selected DRGs in the field of the HDFS
using the criteria of \cite{Fr03} yielding 14 galaxies with photometric redshifts
(Rudnick et al. 2003) in the range $1.9 < z < 3.8$. 
One blended source was excluded. In addition, we selected isolated LBGs 
in the same field to the same $K_s$-band limit and in the same redshift range, 
using the U-dropout criteria of \cite{Ma96}
on the WFPC2 imaging \citep{Ca00}. The two samples are complementary,
with only 1 source in common. 
For both samples, the 
mean redshift is $z=2.6$ with a standard deviation of 0.5\footnote{ 
The photometric redshift accuracy of DRGs from other studies is 
$|z_{spec} - z_{phot}|/(1+z_{spec})\approx0.1$ (F04, Wuyts et al. 
in preparation). The estimate is based on
direct comparison of 16 galaxies with both photometric and spectroscopic 
redshifts. Uncertainties in rest-frame magnitudes 
and model parameters are based on Monte Carlo simulations that take into
account uncertainties in the observed fluxes and redshift estimates.}.

\figa
\figb

\section{Red Galaxies at $z>2$: old or dusty or both?}

In Fig.~\ref{fig.a} we analyze the  $I-K_s$ versus the 
$K_s-4.5\mu$m colors. The $K-4.5\mu$m color has a sufficiently large baseline and much higher
signal-to-noise ratio than the $K_s-5.8\mu$m or $K_s-8\mu$m colors. 
The DRGs and LBGs lie in very different regions, with little overlap.  The mean
$K-4.5\mu$m color of the DRGs is significantly redder than that of LBGs,
confirming that DRGs have higher $M/L$ ratios in the
rest-frame optical.  The red $I-K_s$ and $K-4.5\mu$m colors of the DRGs imply that they must
be prominent in IRAC selected samples. Indeed, the majority of 
red $z-3.6\mu$m sources selected by Yan et al. (2004) in the GOODS survey
satisfy the DRG selection criteria.
We show color tracks for stellar population models
(Bruzual \& Charlot 2003; BC03), redshifted to a fixed $z=2.6$: a
single-age burst model (SSP) and two models with constant star formation (CSF), and
reddenings typical for LBGs ($A_v=0.6$, Shapley et al 2001), and typical
for DRGs ($A_V=2.0$, F04). The $K_s-4.5\mu$m
colors of LBGs are consistent with low reddening models, whereas
most of the DRGs lie in the area of models with substantial
extinction.

\subsection{Dead galaxies at $z=2-3$}

Three of 13 DRGs (indicated by star symbols in Fig.~1) lie well outside the area
of constant star-forming models. Their $I-K_s$ colors are too red for their
$K-4.5\mu$m  color and they lie close to the line of an old single-age
burst model with ages of $2-3$ Gyr.
The candidate old, passive galaxies are shown in 
Fig.~\ref{fig.b}. Their SEDs are very well fit by an 
old single burst population. 
The gray model curves are predictions based on fits to the exceptionally deep 
optical/NIR data only, demonstrating that the Spitzer fluxes lie 
very close to the expected values.
Dead galaxies have blue $K-4.5\mu$m colors and with Spitzer we
can now effectively separate them from dusty star-forming DRGs, which 
are red in $K-4.5\mu$m.

Models with ongoing star-formation and dust-reddening fit the SEDs badly.
Star-formation histories 
(SFHs) with exponentially declining star-formation rates (SFRs) 
and timescales $> 300$ Myr can be ruled out at the 99.9\% confidence level.  
For the marginally acceptable
models, the ratio of on-going to past-average SFR is $SFR(t)/<SFR> = 0.001$
indicating that these 3 galaxies are  truly ``red
and dead''. The best-fit ages and implied formation redshifts depend on
the assumed  metallicity $Z$, as expected.   Super-solar
$Z=0.05$ models give a  mean age of $<t>=1.4$~Gyr and mean formation
redshift $<z_f>=5$, solar $Z=0.02$ models yield $<t>=2.9$ Gyr and $<z_f>
\gg 5$,  while sub-solar metallicity models fail to provide good fits to
the IRAC fluxes as they are too blue. Hence we infer from the models that
the ``dead'' galaxies had formed most  of their stellar mass by $z\sim5$
($Z=0.05$) or much earlier $z\gg5$ ($Z=0.02$).

The number density of ``dead'' galaxies is 1.9 $\times 10^{-4} h_{70}^3$~Mpc$^{-3}$, assuming
a top hat redshift distribution between $z=2$ and 3.
For stellar masses  $> 0.5 \times 10^{11} M_{\sun}$ and the same IMF,
that number density 
is $10\times$ lower than that of early-type galaxies in the nearby universe 
(Bell et al. 2003). Obviously, the sampled volume is too small to allow firm 
conclusions, but the result is indicative of  strong evolution of passive galaxies from $z=2.5$ to $z=0$.

\subsection{Dusty star forming galaxies}

In Fig.~\ref{fig.b} we show the SEDs for 3 of the 8 DRGs
whose optical/NIR fluxes were better fit with CSF models
and reddening by a Calzetti et al. (2000) dust law.
 The observed MIR flux points are 
often different from predictions based on shorter wavelengths, especially where the 
flux density $F_\lambda$ was still rising at the observed 
 $K_s-$band. Hence, for the very dusty DRGs ($A_V>1.5$) the MIR fluxes 
 help to better constrain the age and dust in the models. 
 The average ages and extinction of the fits changed mildly after 
 inclusion of the IRAC data: 
from $<t>=$1.1 to 1.3 Gyr, and $<A_V>=$1.9 to 1.5 mag, respectively.

Finally, two remaining galaxies could not be fit well by any model, 
most likely due to emission line contamination. 
One of them is spectroscopically confirmed as a very strong 
Lyman-$\alpha$ emitter (Wuyts et al. in preparation). However,
the generally good fits to the UV-to-8$\mu$m SEDs of the DRGs
indicate that the red $J_s-K_s$ colors were caused by old age and dust-reddened 
star light, and not other anomalies. We note that one of the dead galaxies 
(see Fig.~1) has an apparent flux excess at $8\mu$m,
suggesting the presence of an obscured AGN which
starts to dominate the flux in the rest-frame $K-$band (see e.g., \citealt{Ste05}).

\section{Mass-to-light ratio variations at $z=2-3$}
We next study the $M/L$ ratios inferred from the SED fits described
above, adopting the best fitting SFH (SSP or CSF).  
Figure 3a shows the modeled $M/L_K$ (rest-frame $K$)  versus
rest-frame $U-V$.  The curves show the dust-free BC03 model $M/L_K$
ratios. Galaxies  generally lie to the red of the model curves as a
result of dust attenuation.  The fits imply a large range in
$M/L_K$  for DRGs and LBGs together (a factor of 6). Furthermore,
the $M/L_K$ ratio correlates with rest-frame $U-V$ color,  
where galaxies red in the rest-frame optical have higher
$M/L_K$ ratios than blue galaxies. 
The average $M/L_K$ of the DRGs
is $0.33 \pm 0.04$, about three times higher than that of the LBGs. 
Figure 3b  shows the derived $M/L_K$ against the derived age. The $M/L_K$
correlates well with the age, as expected from the models. The role
of extinction is greatly reduced in the rest-frame $K-$band, 
implying that the differences in the $M/L_K$ ratios for DRGs and
LBGs are driven by age differences.

Figure 3c shows the $M/L_K$ ratio against stellar mass $M_*$. DRGs dominate the high-mass end.
The highest-mass galaxies ($> 0.5 \times 10^{11} M_{\sun}$) in this sample all have high 
$M/L_K$ ratios, and here the $M/L_K$ ratio does not depend strongly on $M_*$. 
At lower masses galaxies have much lower $M/L_K$ ratios. This may be a 
selection effect caused by our magnitude cut-off, as we would miss 
low-mass galaxies with high $M/L_K$. However, high-mass galaxies 
with {\it low} $M/L_K$ would be detected if they existed, hence the 
lower envelope of the $M/L_K$ versus $M_*$ distribution is real. 
Two intriguing possibilities are that 
the mean $M/L_K$ decreases towards lower stellar mass, or that the intrinsic 
scatter in $M/L_K$ increases.

Our DRGs and LBGs were selected in the observed $K_s-$band (rest-frame $V-$band). 
We find a large variation in $M/L_V$ (a factor of 25), hence
selection effects play a major role in NIR studies at high redshift.
Selection using MIR observations with Spitzer
would improve the situation, but the wide range in $M/L_K$ ratios and ages
found here indicates that even a IRAC $8\mu$m$-$selected sample 
would still be very different from a mass-selected sample. 

\figc 

\section{Discussion and conclusions}
We have presented rest-frame UV-to-NIR photometry of a sample
of DRGs and LBGs at $z=2-3$ in the HDFS. These
galaxies span a wide range in properties, similar to low-redshift 
galaxies. 
The rest-frame NIR photometry from IRAC
helps significantly: first, 
by allowing us to separate ``old
and dead''  from dusty star forming DRGs using only the
observed $I,K_s$, and $4.5\mu$m-band, and second, by  improving model
constraints on the heavily obscured DRGs.

The wide range in galaxy properties at $z=2.5$ raises several
important issues. First, it demonstrates it is impossible to
obtain  mass-selected samples photometrically. Even in the 
rest-frame $K$-band, the $M/L$ ratio varies by a factor
of 6 for the DRGs and LBGs in the HDFS. Second, we need to
understand what produces these variations. If a
relation between total stellar mass and $M/L_K$ exists
(see Fig.~3c) then stellar mass might be driving the variations. 
Deeper IRAC data is needed to establish this well, as 
incompleteness effects may play a role.
It is tantalizing that Kauffmann et al. (2003) find a similar correlation 
at $z=0.1$, with the most  massive galaxies being the  oldest and having the
highest  $M/L$ ratios. The authors also found that above a stellar
mass $M_*=6 \times 10^{10} M_{\sun}$ 
galaxy properties correlate only weakly with $M_*$, 
similar to what we find at $z=2-3$. The simplest explanation is that we 
observe the same galaxies at $z=0.1$ and $z=2.5$, although we note 
that hierarchical formation scenarios predict significant merging 
for galaxies at $z>2$.  Alternatively, we observe merging processes occurring at a
critical  mass of about $6 \times 10^{10} M_{\sun}$.  A partial
explanation may be a simple relation between mass and metallicity:
higher mass galaxies might have higher metallicity, and thus be
redder.

It is very noteworthy that there are ``red and dead'' galaxies at $z=2-3$. 
Previous authors found such galaxies 
at $z=1-2$ \citep[e.g.,][]{Ci04,MC04}. Whereas the
number density of ``dead'' galaxies at $z=2.5$ is probably much lower than
at $z=0$, the mere existence of these systems at such high redshift
raises the question what caused such a rapid and early
decline in star formation. Our model fits imply that their star formation
stopped by $z=5$ or higher, close to or during the epoch of reionization.
Possibly a strong feedback mechanism caused this, such as an AGN or 
galactic-scale outflow. We note that these galaxies are among the most massive 
galaxies at $z=2.5$, and hence were probably at the extreme tail 
of the mass function at $z=5$.

Obviously, many questions remain unanswered by this study. Only a very small
field has been studied, and similar studies on wider fields are necessary.
Finally, the $M/L$ ratios derived here remain model dependent 
and vary with the assumed SFH, metallicity, and IMF (e.g., Bell et al. 2003). 
Direct mass determinations 
are presently very challenging, but are becoming more important to
understand the $z=2-3$ galaxy population.

\begin{acknowledgements}
This research was supported by the Carnegie Institution of Washington, the Netherlands 
Foundation for Research (NWO), the Leids Kerkhoven-Bosscha Fonds, the Lorentz Center, 
and the Smithsonian Institution. 
\end{acknowledgements}


\begin{thebibliography}{}
\bibitem[Barmby et al.(2004)]{Ba04} Barmby, P., et al.\ 2004, \apjs, 154, 97 
\bibitem[Bell et al.(2003)]{Bel03} Bell, E.~F., et al.\ 2003, \apjs, 149, 289 
\bibitem[Bruzual \& Charlot(2003)]{BC03}  Bruzual, G.~\& Charlot, S.\ 2003, \mnras, 344, 1000 (BC03)
\bibitem[Casertano et al.(2000)]{Ca00} Casertano, S. et al. \aj, 120, pp.~2747--2824, 2000
\bibitem[Calzetti et al.(2000)]{Cal00} Calzetti, D., et al. \  2000, \apj, 533, 682 
\bibitem[Cimatti et al.(2004)]{Ci04} Cimatti, A., et al.\ 2004, \nat, 430, 184 
\bibitem[Fazio et al.(2004)]{Fa04} Fazio, G.~G., et al.\  2004, \apjs, 154, 10 
\bibitem[Franx et al.(2003)]{Fr03} Franx, M.~et al.\ 2003, \apjl, 587, L79
\bibitem[F\"orster Schreiber et al.(2004)]{Fo04b} F\"orster Schreiber, N.~M., et al.\ 2004, \apj, 616, 40 (F04) 
\bibitem[Kauffmann et al.(2003)]{Ka03a} Kauffmann, G., et al.\ 2003, \mnras, 341, 54 
\bibitem[Labb{\' e} et al.(2003)]{La03} Labb{\' e}, I., et al.\ 2003, \aj, 125, 1107
\bibitem[Labb{\' e} et al.(2004)]{La04} Labb{\' e}, I., \ 2004, Doctoral Thesis, Leiden University  
\bibitem[Madau et al.(1996)]{Ma96} Madau, P., et al.\ 1996, \mnras, 283, 1388
\bibitem[McCarthy et al.(2004)]{MC04} McCarthy, P.~J., et al.\ 2004, \apjl, 614, L9 
\bibitem[Papovich, Dickinson, \& Ferguson(2001)]{PDF01} Papovich, C., Dickinson, M., \& Ferguson, H.~C.\ 2001, \apj, 559, 620
\bibitem[Rudnick et al.(2003)]{Ru03} Rudnick, G., et al.\ 2003, \apj, 599, 847 
\bibitem[Steidel et al.(1996a)]{St96a} Steidel, C.~C., et al. \ 1996, \aj, 112, 352
\bibitem[Steidel et al.(1996b)]{Ste96b} Steidel, C. C., et al. 1996b, \apj, 462, L17
\bibitem[Stern et al.(2005)]{Ste05} Stern, D., et al. \ 2005, \apjl, submitted (astro-ph/0410523)
\bibitem[van Dokkum et al.(2003)]{vD03} van Dokkum, P.~G.~et al.\ 2003, \apj, 587, L83
\bibitem[van Dokkum et al.(2004)]{vD04} van Dokkum, P.~G., et al.\ 2004, \apj, 611, 703 
\bibitem[Yan et al.(2004)]{2004ApJ...616...63Y} Yan, H., et al.\ 2004, \apj, 616, 63 
\end{thebibliography}
\end{document}